# LOCAL AND FUNDAMENTAL MODE COUPLER DAMPING OF THE TRANSVERSE WAKEFIELD IN THE RDDS1 LINACS

R.M. Jones[1]; SLAC, N.M. Kroll[2]; UCSD & SLAC, R.H. Miller[1], C.-K. Ng[1] and J.W. Wang[1]; SLAC

Abstract
In damping the wakefield generated by an electron beam traversing several thousand X-band linacs in the NLC we utilise a Gaussian frequency distribution of dipole modes to force the modes to deconstructively interfere, supplemented with moderate damping achieved by coupling these modes to four attached manifolds. Most of these modes are adequately damped by the manifolds. However, the modes towards the high frequency end of the lower dipole band are not adequately damped because the last few cells are, due to mechanical fabrication requirements, not coupled to the manifolds. To mitigate this problem in the present RDDS1 design, the output coupler for the accelerating mode has been designed so as to also couple out those dipole modes which reach the output coupler cell. In order to couple out both dipole mode polarizations, the output coupler has four ports. We also report on the results of a study of the benefits which can be achieved by supplementing manifold damping with local damping for a limited number of cells at the downstream end of the structure.

## 1. INTRODUCTION

The transverse wakefield in an accelerator structure is due to dipole modes which are excited in the structure when the beam traverses it off center. In the manifold damped detuned structures [1] two measures are taken which mitigate the deflecting effect of these modes on trailing bunches. The primary measure taken is to modify the individual cells so as to achieve a smooth (currently a kick factor weighted truncated Gaussian) distribution of mode frequencies which via destructive interference between the modes leads to a large decrease in deflecting forces on bunches arriving some 1.4 to 7 nanoseconds later. Due to the discreteness of the modes this effect begins to break down at subsequent time delays, earliest for the most widely spaced modes, latest for the most narrowly spaced (ie where the mode density in frequency is highest). This latter point, typically a maximum, is often referred to as the recoherance peak because it occurs where the most closely spaced modes are again in phase (~100 ns). The damping manifolds, four tapered circular waveguides with electric field coupling to all cells except a few at the ends, which serve to drain amplitude from these modes with a time constant ~20 ns thereby limiting the magnitude of the recoherance effect and ultimately leading to a steady reduction in wake amplitude. This paper is devoted to a discussion of two aspects of this scenario which have limited the effectiveness of this approach. As mentioned above a few cells at the ends are not coupled to the manifolds. This has been done to avoid mechanical interference problems. The consequence has been that one to a few modes at the high frequency end and with significant kick factors are very poorly damped. They show up as sharp peaks in the high frequency end of the spectral function and elevate the wake amplitude. This effect has been more prominant in the RDDS design than in the earlier DDS designs due to the increase in the detuning frequency span (11.25%) and in the ratio of frequency span to Gaussian width (4.75). Nevertheless it was manifold radiation observed in the DDS3 ASSET experiment which suggested a cure. Sharp high frequency peaks were observed in the manifold radiation when the beam was displaced vertically but not when it was displaced horizontally suggesting that these modes were being damped by the output coupler for the accelerating mode which then had two horizontal waveguide ports. This led to the design of a four port output coupler intended to damp both polarizations of the dipole modes.

The design of the coupler, the equivalent circuit analysis of its effect and confirmation by simulation, by RF measurements, and by observation of beam induced radiation from the output coupler will be discussed in following section. The other aspect has to do with the rise in the wake amplitude which occurs at the earliest times and is due to the early onset of recoherance for the most widely spaced modes. The effect of manifold damping does not set in soon enough to mitigate this effect. This led to an investigation of the effect of locally damping a limited number of cells at the downstream end of the structure. The results of this study will follow the discussion of the four port output coupler

## 2. THE FOUR PORT OUTPUT COUPLER

*(1) Design*

The standard output coupler for the DDS structures has been fitted with a pair of WR90 rectangular output waveguides polarized in the electric field direction emerging on opposite sides of the coupler cell in the horizontal direction. In order to provide an output for both dipole mode polarizations these output waveguides have been replaced by four WR62 waveguides oriented so as to form an "X" with respect to the horizontal and vertical axes [1]. The reduced width was chosen to conveniently fit the space limitations dictated by the output cell dimensions, and the orientation was chosen to avoid mechanical interference with the manifold couplers, which are located in the vertical and horizontal planes. The accelerating mode and the lower dipole modes, while the $TE_{01}$ polarisation is cut-off for both of them. The coupler was matched for the accelerator mode by varying the cell radius and wavguide iris width as described in [2] using the mesh shown in Fig. (1).

[1]Supported under U.S. DOE contract DE-AC03-76SF00515.
[2]Supported under U.S. DOE grant DE-FG03-93ER407.

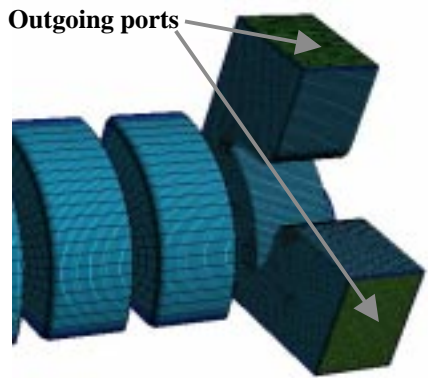

Fig 1: Fundamental four-port coupler

## *(2) Assessment of the damping effect*

The first step in the assessment was the determination of the loading effect on the output cell. A symmetry reduced section of the output cell with dipole π mode boundary conditions was probe driven in a broad band time domain simulation with outgoing wave boundary conditions on the output waveguides to determine the resonant frequency and $Q_{ext}$. The KY [3] frequency domain method was also applied for corroboration, and a $Q_{ext}$ of 36 established. This information was used in a number of

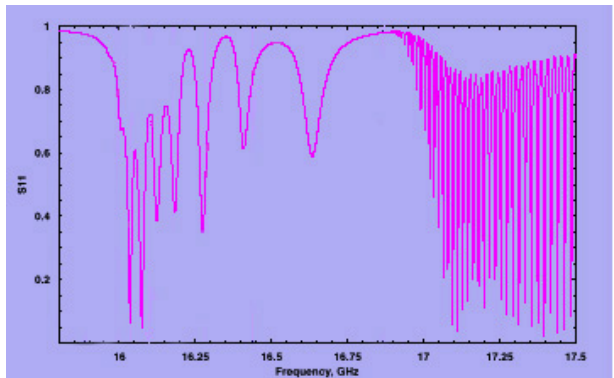

Fig. 2: Circuit model of $S_{11}$ of output port

ways in the equivalent circuit calculations to investigate the damping effect on the structure modes. An effect is expected of course only for those modes which reach the output cell, and all of the Q's are expected to be much higher than the 36 referenced above. The most direct equivalent circuit information was obtained by computing the dipole mode $S_{11}$ as seen from the output coupler, parameterised by making use of the simulation determined $Q_{ext}$ (Fig. (2)). The $Q_{ext}$ of the broadest peak shown was of the order of 300, a figure which agrees with the result obtained from a time domain simulation of a ten-cell model (without manifolds) terminated with the output coupler. One notes that the peaks narrow as one moves to the lower frequency peaks due to an increase in $Q_{ext}$. This effect combined with the onset of manifold damping leads to the total disappearance of resonance effects below 16 GHz. The spectral function was recomputed with an appropriate series resistance inserted in the TM circuit of the last cell [4]. The sharp peaks which had been observed previously

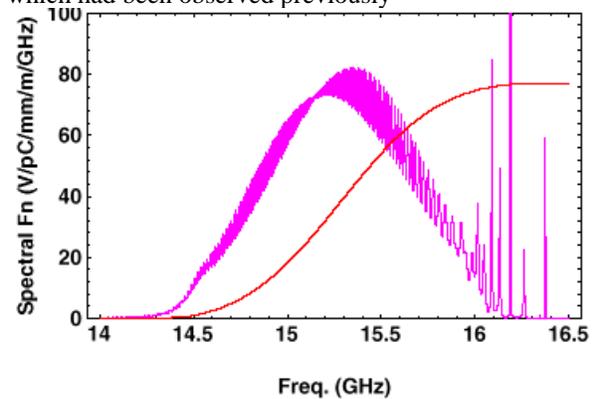

Fig. 3 Spectral function RDDS1 for 4 cells decoupled either end of structure and no external loading of cells

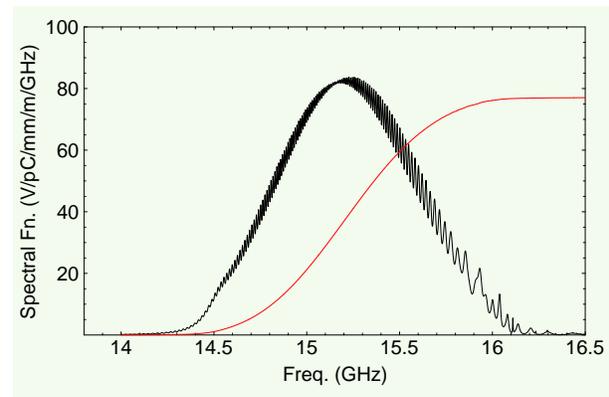

Fig. 4: Spectral function for 4 cells decoupled either end of structure and Q ~36 for the last cell.

(Fig. (3)) were no longer observed (Fig. (4)) and Q's estimated by Lorentzian based fits to spectral function amplitudes were in qualitative agreement with those observed as described above. The associated wake envelope functions are constructed in Figs (5) and (6), and a striking improvement may be noted

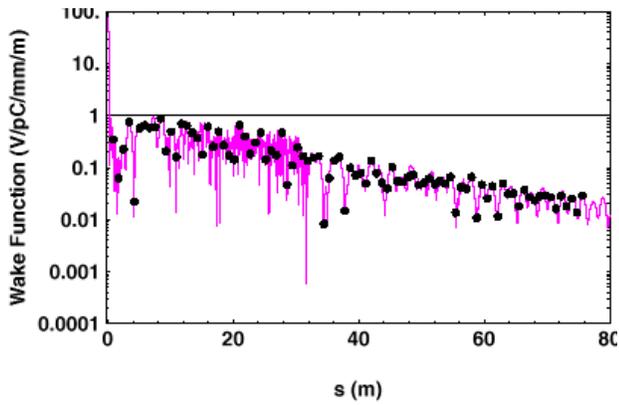

Figure 5: Envelope of wake function for RDDS1 excluding external loading

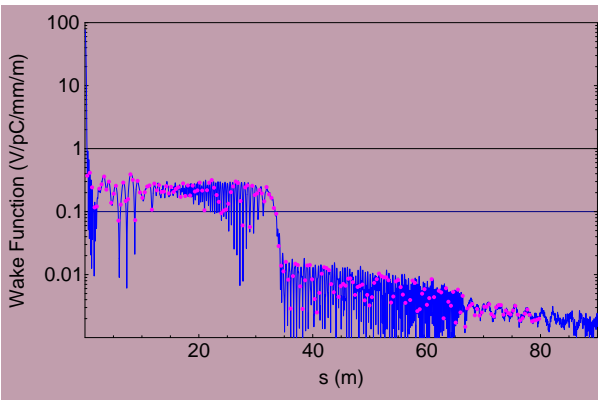

Fig. 6: Envelope of wakefield for 4 cells decoupled either end of structure and Q~36 for the last cell

## III) The Effect of Damping a Limited Number of Cells

The spectral function plotted in Fig. (4) shows a number of peaks on the downward sloping high frequency end which are more widely spaced and somewhat stronger than those found at the lower frequencies. As noted above, because of their wide spacing their contribution to deconstructive interference disappears at quite early times and is responsible for the sharp rise in the wake which follows the first minimum. This observation suggested an exploration of the effect of adding moderate local damping to a limited number of cells towards the output end of the structure. The investigation has so far been completely phenomenological, that is, based on an assumed distribution of Q values rather than a design of the damped cells. We have explored tapered distributions of Q's confined to the cells with cell numbers above 187 and values no less than 500. The best results were obtained with distributions which smoothed out the peaks while leaving the mean value of the spectral function more or less intact. Figure (7) shows a spectral function resulting from a linear taper of 1/Q from 1/6500 at cell

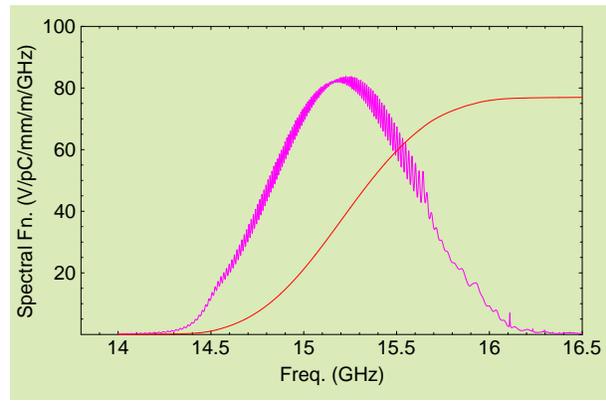

Fig 7: Spectral function for RDDS with the last 10 Cells Q =500 and 1/x taper up to 6500 from Cells 197 to 187.

187 to 1/500 at cell 197 and thence constant. The ripples are indeed smoothed out and the Gaussian like form is intact. A strong suppression of the early rise in the wake may be noted Fir. (8). These results suggest that this approach warrants further investigation.

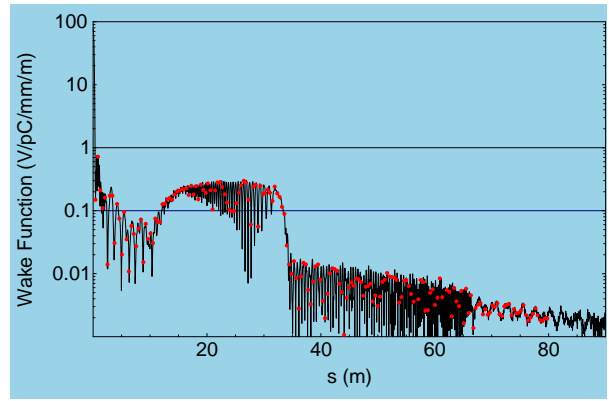

Fig 8: Wake envelope function for RDDS with the last 10 Cells Q =500 and 1/x taper up to 6500 from Cells 197 to 187.

## 4. CONCLUSIONS

We conclude that the four port output coupler undoes the harmful effects of decoupling the last few cells from the damping manifolds. Alternatively this might be done by locally damping a limited number of cells at the output end of the structure with additional benefits, especially in the early time portion of the wake.

## 5. REFERENCES

[1] J.W. Wang et al, TUA03, LINAC200 (this conf.)
[2] N.M. Kroll et al, TUE04, LINAC2000, (this conf.)
[3] N.M. Kroll and D.U.L. Yu, Part. Acc.,34, 231 (1990)
[4] R.M. Jones, et al, LINAC96, (also SLAC-PUB-7287)